\documentclass[preprint,12pt]{elsarticle}
\usepackage{hyperref,amsmath,amsthm,amsfonts}
\DeclareMathOperator{\img}{Im}
\usepackage{amssymb}
\usepackage{amsthm}

\journal{Physics Letters B}

\biboptions{sort&compress}
\begin{document}

\begin{frontmatter}



\title{Classical and Quantum Cosmology of Fab Four John Theories}


%\author[a]{Isaac Torres}
%\ead[a]{itsufpa@gmail.com}
%\address[a]{PPGFis, CCE - Universidade Federal do Esp\'{i}rito Santo, \\ zip 29075-910, %Vit\'{o}ria, ES, Brazil}

\author[a]{Isaac Torres\corref{cor1}}
\ead{itsufpa@gmail.com}
\author[b,c]{J\'{u}lio C. Fabris}
\author[b,d]{Oliver F. Piattella}
%\ead{oliver.piattella@cosmo-ufes.org}

\cortext[cor1]{Corresponding author}
\address[a]{PPGFis, CCE - Universidade Federal do Esp\'{i}rito Santo, zip 29075-910, Vit\'{o}ria, ES, Brazil}
\address[b]{N\'{u}cleo Cosmo-ufes \& Departamento de F\'{i}sica - Universidade Federal do Esp\'{i}rito Santo zip 29075-910, Vit\'{o}ria, ES, Brazil}
\address[c]{National Research Nuclear University MEPhI, Kashirskoe sh. 31, Moscow 115409, Russia}
\address[d]{Institut f\"ur Theoretische Physik, Ruprecht-Karls-Universit\"at Heidelberg, Philosophenweg 16, 69120 Heidelberg, Germany}

%Declarations of interest: none.

\begin{abstract}
We study the John term of Fab Four cosmology in the presence of a scalar potential. We show here how this theory can describe a wide range of cosmological solutions. This theory has two general functions of the scalar field: the potential $V(\phi)$ and the John coefficient function $V_j(\phi)$. We show that for very simple choices of those functions, we can describe an accelerated expansion, a radiation-dominated era, and a matter-dominated era. By means of simple modifications, it is also possible to describe nonsingular bouncing versions of those solutions and cyclic universes. We also address some quantum issues of that theory, showing that, for the most significant singular cases, the theory admits a classically well behaved quantization, even though the Hamiltonian has fractional powers in the momenta.
\end{abstract}


\end{frontmatter}
\section{Introduction}
In \cite{PhysRevD.80.103505,PhysRevD.85.123520,PhysRevD.88.083539}, it is described how a cosmological theory with a scalar field nonminimally coupled to gravity, represented by the Lagrangian density
\begin{equation}\label{rpg5mv}
L=\sqrt{-g}\left[\frac{R}{8\pi}-\nabla_{\mu}\phi\nabla^{\mu}\phi-\kappa G^{\mu\nu}\nabla_{\mu}\phi\nabla_{\nu}\phi -V(\phi)\right],
\end{equation}
for $V=0$, drives an inflationary epoch, followed by a ``graceful'' exit from inflation, thanks to the presence of the nonminimal coupling term $G^{\mu\nu}\nabla_{\mu}\phi\nabla_{\nu}\phi$. Theories similar to \ref{rpg5mv} can be found in \cite{PhysRevLett.105.011302,Starobinsky_2016,Bruneton:2012zk}. In \ref{rpg5mv}, $R$ is the Ricci scalar, $\phi$ is a scalar field, $G^{\mu\nu}$ is the Einstein tensor, and $\kappa$ is a coupling constant.

The theory \ref{rpg5mv} is a subclass of Horndeski modified gravity \cite{Horndeski:1974wa}, the most general scalar-tensor gravitational theory in four dimensions with second order equations of motion. Modifying gravity is an alternative to general relativity to explain observations of the accelerated expansion of the Universe \cite{Capozziello:2010zz,Papantonopoulos:2015cva}. In this sense, Horndeski theory is particularly important, since it is a modification of gravity that avoids Ostrogradsky instability \cite{Ostrogradsky:1850fid,Kobayashi_2019} and includes the general theory of relativity as a particular case. The Horndeski action is written as
\begin{equation}\label{horndact}
	S_H=\int d^4x\sqrt{-g}(L_2+L_3+L_4+L_5),
\end{equation}
where
\begin{align}\label{hrndskip}
	L_2 & =K(\phi,X),  \\
	L_3 & =-G_3(\phi,X)\Box\phi,  \\
	L_4 & = G_4(\phi,X)R+G_{4,X}(\phi,X)[(\Box\phi)^2  \nonumber\\ &-\nabla^{\mu}\nabla^{\nu}\phi\nabla_{\mu}\nabla_{\nu}\phi],\\
	L_5 & = G_5(\phi,X)G^{\mu\nu}\nabla_{\mu}\nabla_{\nu}\phi-\textstyle\frac{1}{6}G_{5,X}(\phi,X) [(\Box\phi)^3\nonumber\\
	& -3\Box\phi\nabla^{\mu}\nabla^{\nu}\phi\nabla_{\mu}\nabla_{\nu}\phi   \nonumber\\ &+2\nabla_{\mu}\nabla_{\nu}\phi\nabla_{\lambda}\nabla^{\mu}\phi\nabla^{\nu}\nabla^{\lambda}\phi].
\end{align}
The functions $K$ and $G_i$ are generic differentiable functions of the scalar field $\phi$ and of the kinetic term $X\equiv-\nabla^{\mu}\phi\nabla_{\mu}\phi$. The notation $G_{i,X}$ denotes the derivative of $G_{i}$ with respect to $X$. The greek indices here run from $0$ to $3$.

Another application of the nonminimal coupling term $G^{\mu\nu}\nabla_{\mu}\phi\nabla_{\nu}\phi$ for accelerated expansion comes from the Fab Four theory, which is the most general subclass of the Horndeski theory with a self-tuning mechanism able to deal with the cosmological constant problem \cite{PhysRevLett.108.051101}. In \cite{PhysRevLett.108.051101,PhysRevD.85.104040,1475-7516-2012-12-026}, it is shown how the non-minimal coupling represented by the so-called ``John'' Lagrangian,
\begin{equation}\label{lj}
L_{john}=V_j(\phi)G^{\mu\nu}\nabla_{\mu}\phi\nabla_{\nu}\phi,
\end{equation}
 helps the other three Fab Four terms to provide all the usual epochs of cosmic evolution, in the presence of a matter action. In \ref{lj}, $V_j(\phi)$ is a free function of the scalar field, related with the coefficient function $G_5$ of Horndeski theory (see \cite{Capozziello:2010zz}, for instance) by $V_j=\partial G_5/\partial \phi$. If considered alone, without any potential, $L_{john}$ represents a stiff matter-dominated universe, when the spatial curvature is subdominant \cite{1475-7516-2012-12-026}. Another application of $L_{john}$ is for Galileon black holes \cite{Babichev2014}. Of course, those are just a few examples.
  
 The above mentioned applications usually set the coefficient function $V_j$ from the start and consider \ref{lj} only as a contribution for the Lagrangian of a minimally coupled scalar field, like in \ref{rpg5mv}. Therefore, the specific dynamics of \ref{lj} for a general $V_j$ function has not been studied yet. In this paper, we are interested in the theory represented by\\
 \begin{equation}\label{lagtot}
 L=\sqrt{-g}\left[-V_j(\phi)G^{\mu\nu}\nabla_{\mu}\phi\nabla_{\nu}\phi -V(\phi)\right],
 \end{equation}
 where the potential was introduced to avoid trivial solutions. Throughout all this letter, we will  consider the spatially flat Friedmann-Lema\^itre-Robertson-Walker metric
 \begin{equation}\label{metfrwn}
 ds^2 = N^2dt^2 - a^2\delta_{ij}dx^idx^j,
 \end{equation}
 where $N(t)$ is the lapse function \cite{Calcagni:2017sdq} and $a(t)$ is the scale factor. Observe that \ref{lagtot} is still a subclass of Horndeski modified gravity \ref{horndact}, for $K(\phi,X)=V(\phi)$.
  
% The recent detection of the gravitational waves imposed a constraint over models containing the nonminimal coupling $L_{john}$ \cite{Kobayashi_2019}. Howerver, that constraint does not excludes $L_{john}$. Instead, it says that $G_5$ must be small in comparison with, let us say, the Einstein-Hilbert term for redshift $z\leq0.009$ \cite{Kase:2018aps}. Thus, the theory \ref{lagtot} must be seen as an effective contribution that can still be studied. In this letter, our purpose is to discuss which is the effect of \ref{lagtot} if taken alone. Thus, we investigate here what is the specific dynamics of $L_{john}$ to have a better understanding of its effects when it is coupled with another theory, like \ref{rpg5mv}.
 
 %\subsection{Parte nova}
%{\bf Referências novas:}
%\begin{enumerate}
%	\item \cite{Felice_2012}: Perturbações em Horndeski, 2012.
%	\item \cite{Bellini_2014}: Artigo que introduziu os $\alpha_i$'s, 2014.
%	\item \cite{PhysRevLett.119.161101}: evento GW170817, 2017.
%	\item \cite{Goldstein_2017}: evento GRB 170817A, 2017
%	\item \cite{Abbott_2017}: Vínculo de $c_{gw}$, 2017
%	\item  \cite{Kase:2018aps}: Como GW vinculou Horndeski, 2019.
%	\item \cite{Gong2018}: Apesar do vínculo, é possível usar $G_5$ sem fine tuning, 2018.
%\end{enumerate}

The recent observational events GW170817 and GRB 170817A have imposed the constraint \cite{PhysRevLett.119.161101,Goldstein_2017,Abbott_2017}:
\begin{equation}\label{vinculogw}
-3\times10^{-15}\leq\frac{v_{\text{GW}}-v_{\text{EM}}}{v_{\text{EM}}}\leq+7\times10^{-16},
\end{equation}
where $v_{\text{GW}}$ is the speed of gravity and $v_{\text{EM}}$ is the speed of light. Thus, it became possible to test alternative theories of gravity for tensor perturbations. In \cite{Felice_2012}, it was described how linear perturbations can be performed in the general Horndeski theory and in \cite{Bellini_2014} a complete set of parameters were introduced to simplify the comparison between theory and observations for those perturbations. Some authors (for instance, \cite{Kobayashi_2019,Kase:2018aps}) argue that this constraint completely rule out some Horndeski theories (like \ref{rpg5mv}) to avoid fine tuning. But some other authors argue the opposite, for the following reasons. First,  \ref{vinculogw} restricts $v_{\text{GW}}$ only for the low redshit range $z\lesssim0.01$, as far as we know \cite{Abbott_2017,PhysRevD.98.044051}. In other words, $G_5$ may be relevant in the early universe, in accordance with previous works \cite{PhysRevD.80.103505,PhysRevD.85.123520,PhysRevD.88.083539}. Second, for  \ref{rpg5mv}, the derivative coupling  $G^{\mu\nu}\nabla_{\mu}\phi\nabla_{\nu}\phi\sim H^2\dot{\phi}^2$, where $H$ is the Hubble parameter, decreases as the universe expands, thus becoming negligible in comparison with the kinetic term $\sim\dot{\phi}^2$. Hence, for the redshift values for which \ref{vinculogw} is valid, we expect that the derivative coupling generates only a tiny variation of $v_{\text{GW}}$ from unity \cite{Gong2018}. Third, it is shown in \cite{Gong2018} that the range of values of the mass scale $M$ (roughly speaking, the inverse coefficient) of $G^{\mu\nu}\nabla_{\mu}\phi\nabla_{\nu}\phi$ for which \ref{vinculogw} is valid is
\begin{equation}\label{vincgwhdskg5}
2\times10^{-35}\text{GeV}\lesssim M\ll 10^{15}\text{GeV}.
\end{equation}
Thus, there is no fine tuning in the nonminimal derivative coupling.

In summary, we can say that theories with a $G_5$ term are not ruled out, provided that \ref{vincgwhdskg5} is verified. All the above discussion motivates us to investigate what is the specific cosmology of the derivative coupling alone, in order to have a better understanding of its effects. Strictly speaking, for a more complete description, we should consider \ref{lagtot} as a part of a more general framework, like in \ref{rpg5mv}, but our goal here is precisely to explore the specificity of \ref{lagtot} due to a possible predominant role it can play in the primordial universe. Thus, we will investigate the background cosmology of \ref{lagtot}, which is a minimal non trivial theory containing $G_5$. For the above reasons, we shall focus on the primordial universe, when such a theory can be more effectively relevant. 

%\subsection{Continuação da Intro}
In Section \ref{secham}, we start from \ref{lagtot} with general $V(\phi)$ and $V_j(\phi)$, for a homogeneous scalar field $\phi$, showing that the second order equations of motion can be integrated to become a first order system of equations, for any $V$ and $V_j$. That system has two immediate implications. First, the scalar field must be a time scale, because it is diffeomorphic to cosmic time. Second, that system is a mechanism to provide almost any desired functional form of the scale factor, if suitable $V,V_j$ are chosen. This tuning mechanism can be considered analogous to Fab Four's self-tuning, even though they are different.

Those results for background cosmology show that the non minimal coupling \ref{lagtot} actually covers a wide range of possibilities. That freedom comes from the generality of the coefficient function $V_j$ and the potential $V$. As we will see in Section \ref{secham}, the theory \ref{lagtot} can describe solutions analogous to perfect fluid ones, such as radiation-domination and matter-domination. Those solutions are found when $V$ and $V_j$ are some power law functions. We then show how a de Sitter solution can be obtained, for exponential $V$ and $V_j$. That is a first indication that \ref{lagtot} may be able to describe an inflationary phase. However, the other conditions for inflation need further investigations. Those solutions are all singular, but we will show how the functions $V,V_j$ can be slightly modified, thus causing the singularity to be replaced by a bounce. We also briefly exhibit a simple cyclic universe solution. Therefore, Fab Four John \ref{lagtot} may in principle be an alternative to describe the basic eras of cosmological evolution at the background level, also avoiding singularities. The remaining open questions about the viability of this theory will be investigated in future works.

We also present a first quantum approach to \ref{lagtot}. In Section  \ref{qvctrsec}, we sketch a quantization for power law $V,V_j$, with Bohmian interpretation of quantum mechanics \cite{PhysRev.85.166,PhysRev.85.180,holland_1993}, that can be trivially generalized for the case when $V,V_j$ are both exponential functions. We apply that interpretation because some authors \cite{ACACIODEBARROS1998229,PintoNeto:2004uf,10.2307/193027} argue that standard quantum mechanics should not be applied to primordial universe. In brief, they say that classical exterior domain hypothesis (an implicit assumption of standard interpretation related with measurement \cite{Omnes}) becomes a problem when the system under consideration is the whole universe. In this sense, new approaches to quantum cosmology have been developed with alternative interpretations, particularly with Bohmian quantum mechanics \cite{PhysRev.85.166,PhysRev.85.180}. In \cite{PintoNeto:2004uf,0264-9381-30-14-143001} it is shown that Bohmian interpretation is an alternative for quantum cosmology in various situations, because it avoids the conceptual measurement issue and the problem of time. In practice, in that theory, time is recovered by the guidance equations and the measurement problem is avoided because the same guidance equations provide a way to calculate deterministic solutions. For a review of conceptual problems in quantum cosmology and quantum gravity, see \cite{Kiefer:2013jqa}. In our case, the preliminary quantum result we found is that a consistent Bohmian quantization can be applied to \ref{lagtot}, at least when $V,V_j$ are power law functions. That is a consequence of the existence of a quantum potential of order $\sim\hbar^2$ that vanishes when classical solutions are recovered. Finally, in section \ref{conc} we make some remarks as conclusion.

\section{Classical Cosmology of Fab Four John}\label{secham}
Taking the usual connection satisfying $\nabla_{\alpha}g_{\mu\nu}=0$, for \ref{metfrwn}, we obtain the Ricci tensor components
\begin{subequations}
	\label{tensricci}
\begin{align}
R_{00} &=3\frac{\dot{a}}{a}\frac{\dot{N}}{N}-3\frac{\ddot{a}}{a},\\
R_{0i} &=0, \\
R_{ij} &=\delta_{ij}\frac{a^2}{N^2}\bigg(2\frac{\dot{a}^2}{a^2}+\frac{\ddot{a}}{a} -\frac{\dot{a}}{a}\frac{\dot{N}}{N}\bigg),
\end{align}
\end{subequations}
where the dot denotes derivation with respect to the time $t$. Since the scalar field $\phi=\phi(t)$ is homogeneous, it follows from $g=\det(g_{\mu\nu})=-N^2a^6$ and from \ref{tensricci} that the Lagrangian \ref{lagtot} is written in the minisuperspace as follows:
\begin{align}
L=-3aV_j(\phi)\frac{\dot{a}^2\dot{\phi}^2}{N^3}-Na^3V(\phi). \label{ljmse}
\end{align}
Then, the Euler-Lagrange equations for $N,a,\phi$ can be written as:
\begin{subequations}
	\label{eqelc1}
	\begin{align}
  \frac{\dot{a}^2\dot{\phi}^2}{a^2}-\frac{V}{9V_j} &=0, \label{eqclassn}\\
\frac{\ddot{a}}{a}+2\frac{\dot{a}^2}{a^2}-\frac{V'\dot{a}\dot{\phi}}{Va}&=0, \label{eq2ordcla}\\
\ddot{\phi}-3\frac{\dot{a}\dot{\phi}}{a}+\frac{\dot{\phi}^2}{2} \left(\frac{V_{j}'}{V_{j}}+\frac{V'}{V}\right)&=0, \label{eq2ordclf}
	\end{align}
\end{subequations}
where $f'\equiv df/d\phi$, and we have chosen the cosmic time coordinate by fixing $N = 1$ after deriving the equations. Equation \ref{eqclassn} is a constraint over $\dot{a}$ and $\dot{\phi}$, \ref{eq2ordcla} is the cosmological acceleration equation, and \ref{eq2ordclf} is a Klein-Gordon like equation, describing the dynamics of the scalar field $\phi$. Equation \ref{eqclassn} also impose a condition over $V$ and $V_j$: they must always have the same sign in order to avoid imaginary solutions for $a$ and $\phi$. Defining $\alpha\equiv\ln a$, system \ref{eqelc1} becomes:
\begin{subequations}
	\label{eqelc2}
\begin{align}
   \dot{\alpha}^2\dot{\phi}^2-V/9V_j &=0, \label{eq2classn}\\
   \ddot{\alpha}+3\dot{\alpha}^2-\dot{\alpha}(\ln V)\dot{\:}&=0, \label{eq2ordclas}\\
   \ddot{\phi}-3\dot{\alpha}\dot{\phi}+\textstyle\frac{1}{2}\dot{\phi}[\ln(V_jV)]\dot{\:}&=0. \label{eq2ordclfs}
\end{align}
\end{subequations}
Dividing \ref{eq2ordclas} by $\dot{\alpha}$ and \ref{eq2ordclfs} by $\dot{\phi}$, and then integrating, \ref{eqelc2} becomes a first order system:
\begin{subequations}
	\label{eqelc3}
\begin{align}
\dot{\alpha}&=e^{-3\alpha}V(\phi),  \label{apc}\\
\dot{\phi}&=\textstyle\frac{1}{3}e^{3\alpha}[V(\phi)V_j(\phi)]^{-1/2},  \label{fipc}
\end{align}
\end{subequations}
where the factor $1/3$ comes from the constraint \ref{eq2classn}. Notice that the systems \ref{eqelc2} and \ref{eqelc3} are equivalent, for any $V,V_j$, up to an integration constant that we have set to unity.

It follows from equation \ref{fipc} that the scalar field $\phi$ necessarily represents a time scale, if the product $VV_j$ never vanishes. In mathematical terms, if $V(\phi)V_j(\phi)>0$ for all values of $\phi$, then the right-hand side of \ref{fipc} is always positive, which implies that $\phi$ is a monotonic increasing real function defined on real line. It thus follows from a well-known theorem of real analysis that $\phi(t)$ is a diffeomorphism. In other words, a time scale. Hence, we can restrict the discussion to the simplest possible interpretation $\phi(t)=t$, which is true up to a diffeomorphism. In the following, we show the basic singular, bouncing and cyclic solutions obtained from \ref{eqelc3}.

\subsection{Singular Universes}
Taking
\begin{align}
V(\phi) &=V_0\phi^{\frac{1-w}{1+w}},\label{vvjfp}\\
V_j(\phi) &=\frac{V_0}{4}(1+w)^2\phi^{\frac{3+w}{1+w}},\label{vvjfp2}
\end{align}
where $w$ and $V_{0}$ are positive real constants, we obtain the following power law solutions
\begin{equation}\label{afifp}
a(t)=(t/t_0)^{\frac{2}{3(1+w)}},
\end{equation}
where $\phi(t)=t$ and $t_0=[2/3V_0(1+w)]^{\frac{1+w}{2}}$. The constant $w$ is analogous to the equation of state parameter, at least regarding time evolution of scale factor. For, if $w=1$ the universe is stiff matter-dominated, if $w=1/3$ the universe is dominated by radiation, and if $w=0$, the universe is dominated by dust. Setting now
\begin{align}
V(\phi) &=V_0e^{3\gamma\phi},\label{vvjds}\\
V_j(\phi) &=\frac{V_0}{9\gamma^2}e^{3\gamma\phi},\label{vvjds2}
\end{align}
we can also obtain a de Sitter solutions
\begin{equation}\label{afids}
a(t)=a_0e^{\gamma t},
\end{equation}
where $\gamma$, and $V_0$ are positive real constants, $a_0=(V_0/\gamma)^{1/3}$, and again $\phi(t)=t$ . This shows that John Lagrangian with a scalar potential is able to give basic background cosmological solutions. Notice that \ref{afifp} and \ref{afids} are singular solutions. For \ref{afifp}, the singularity is at $t=0$; for \ref{afids} there is an asymptotic singularity for $t\rightarrow-\infty$.

\subsection{Bouncing Universes}
The power laws for $V,V_j$ found above can be modified to give a nonsingular solution. For
\begin{align}
V(\phi) &=V_0\phi(\phi_{0}^{2}+\phi^2)^{\frac{-w}{1+w}},\label{vvjbfp}\\
V_j(\phi) &=\frac{V_0}{4\phi}(1+w)^2(\phi_{0}^{2}+\phi^2)^{\frac{2+w}{1+w}}, \label{vvjbfp2}
\end{align}
we obtain
\begin{equation}\label{bfp}
a(t)=a_0\left[1+(t/\phi_0)^2\right]^{\frac{1}{3(1+w)}},
\end{equation}
where $\phi(t)=t$. The quantities $\phi_0,V_0$ are positive constants and $a_0= [3V_0(1+w)\phi_{0}^{2/(1+w)}/2]^{1/3}$. The solution \ref{bfp} is a correction to \ref{afifp}, because for $t\gg\phi_0$, they are the same, but for $t=0$, scale factor \ref{afifp} is singular, while \ref{bfp} represents a bouncing universe with minimum radius $a_0>0$. Bounces are an important class of nonsingular cosmological solutions. For a review about bounces in cosmology, see \cite{NOVELLO2008127}. The potentials \ref{vvjbfp} are consistent with \ref{vvjfp}, for $t\gg\phi_0$. The de Sitter solution above can also be replaced by a bouncing, avoiding the singularity at $t\rightarrow-\infty$. Choosing
\begin{align}
V(\phi) &=V_0\sinh(\gamma\phi)\cosh^2(\gamma\phi),\label{vvjbds}\\
V_j(\phi) &=\frac{V_0\cosh^4(\gamma\phi)}{9\gamma^2 \sinh(\gamma \phi)},\label{vvjbds2}
\end{align}
we find
\begin{equation}\label{bdsafi}
a(t)=a_0\cosh(\gamma t),
\end{equation}
where $V_0,\gamma$ are positive constants,  $\phi(t)=t$, and $a_0=(V_0/\gamma)^{1/3}$. In fact, this scale factor represents a bouncing universe that for large values of $t$ reduces to \ref{afids}. Note that the potentials \ref{vvjbds} also reduce to \ref{vvjds}, for large values of $\phi$.

\subsection{Cyclic Universes}
There are some cosmological theories that predict a cyclic universe to avoid initial singularity (see for example \cite{PhysRevD.65.126003}). For \ref{lagtot} it is also possible to obtain such type of solution. Taking, for example,
\begin{align}
	V(\phi) &=V_0\sin(\omega\phi)\left\{a_m+\textstyle\frac{V_0}{\omega}[1-\cos(\omega\phi)]\right\}^2,\\
	V_j(\phi) &=\frac{\left\{a_m+\textstyle\frac{V_0}{\omega}[1-\cos(\omega\phi)]\right\}^4}{9V_0\sin(\omega\phi)},
\end{align}
we obtain the oscillating scale factor
\begin{equation}
	a(t)=a_m+A[1-\cos(\omega t)],
\end{equation}
where $V_0>0$, $A=V_0/\omega$ is the amplitude of oscillation, $\omega$ is the frequency, and $a_m$ is the minimum value of $a(t)$.

\section{Quantum Cosmology of Fab Four John}\label{qvctrsec}
In this section, we briefly show the Hamiltonian formulation of \ref{lagtot}, that has the fractional power $2/3$ in the momenta. Since canonical quantization replaces the momentum by a derivative, the momentum would thus become a fractional derivative.  But there are, in fact, several definitions of fractional derivatives \cite{Herrmann:2011zza}. To avoid this ambiguity, we perform a canonical transformation.  In this first quantum approach, we will consider only the case in which both $V$ and $V_j$ are power law functions of $\phi$. Then, after a short review of basic principles of Bohmian quantum mechanics, we apply a Bohmian quantization to the transformed Hamiltonian. We conclude this section showing the equivalence between classical and quantum equations, for a null quantum potential. This result is expected, because it is the first step to construct a Bohmian quantization. The generalization for the case in which both $V$ and $V_j$ are exponentials follows from the redefinition of the scalar field $\varphi\equiv e^{\phi}$, since for $\varphi$ the Hamiltonian reduces to the former case. Physically, this means that the quantum theory below makes sense for the de Sitter, the radiation-dominated, and the matter-dominated solutions. In this first quantum approach, we will not analyse the nonsingular solutions above, because the canonical transformation described in subsection \ref{cantransfsubsec} imposes a technical restriction.

\subsection{Hamiltonian}
The Hamiltonian follows from the usual Legendre transformation $H(q,p)=\sum\dot{q}_i(q,p)p_i-L(q,p)$, where $q=(N,a,\phi)$ are the generalized coordinates and $p=(p_N,p_a,p_{\phi})$ are the conjugated momenta. It follows from the definition of the momenta that
\begin{subequations}
	\label{afipt}
\begin{align}
\dot{a}=-N(6aV_j)^{-1/3}p_{a}^{-1/3}p_{\phi}^{2/3}, \label{apt}\\
\dot{\phi}=-N(6aV_j)^{-1/3}p_{a}^{2/3}p_{\phi}^{-1/3}. \label{fipt}
\end{align}
\end{subequations}
Therefore, the Hamiltonian is
\begin{equation}\label{hamprobupiaf}
H=N\left[\frac{-3p_{a}^{2/3}p_{\phi}^{2/3}}{2\sqrt[3]{6aV_j(\phi)}} +a^3V(\phi)\right]\equiv N\mathcal{H}.
\end{equation}
Since $p_N\equiv\partial L/\partial\dot{N}=0$, it follows from Hamilton equation $\dot{p}_N=-\partial H/\partial N$ the constraint below:
\begin{equation}
p_{a}^{2/3}p_{\phi}^{2/3}=\frac{2}{3}a^3V[6aV_j(\phi)]^{1/3}. \label{constr}
\end{equation}
From \ref{constr}, it follows also that we can rewrite \ref{afipt} as
\begin{subequations}
	\label{afip}
\begin{align}
\dot{a}=-\frac{2Na^3V}{3p_a},\label{ap}\\
 \dot{\phi}=-\frac{2Na^3V}{3p_{\phi}}. \label{fip}
\end{align}
\end{subequations}
The system \ref{afip} will play a fundamental role in Bohmian quantization, as we shall see next.

\subsection{Canonical Transformation}\label{cantransfsubsec}
The generating function
\begin{equation}\label{funcger}
F(q,P,t)=-\rho a^lP_{x}^{m}-\phi^rP_{y}^{n}+NP_z
\end{equation}
defines a canonical transformation by \cite{goldstein2002classical}:
\begin{subequations}
    \label{canctransf}
\begin{align}
p_i &=\frac{\partial F}{\partial q_i}, \label{tcpj}\\
Q_i &=\frac{\partial F}{\partial P_i}, \label{tcqj}\\
\tilde{H}(Q,P,t) &= H(q,p,t)+\frac{\partial F}{\partial t}, \label{tckh}
\end{align}
\end{subequations}
where $Q=(x,y,z)$ and $P=(P_x,P_y,P_z)$ are the new coordinates and momenta, respectively, and $\tilde{H}$ is the transformed Hamiltonian. The powers $r,l,m,n\in\mathbb{R}-\{0,1\}$ will be fixed later, as well as the positive constant $\rho$. The canonical transformation thus defined is quite restrictive, because the old coordinates become a mix of new coordinates and momenta. Hence we will restrict the discussion for power law $V$ and $V_j$:
\begin{equation}\label{potlptc}
V(\phi)=V_0\phi^{\varepsilon},\qquad V_j(\phi)=V_{j0}\phi^{\delta}.
\end{equation}
It thus follows from \ref{canctransf} that
\begin{align}
\tilde{H}=z\bigg[ -fP_{x}^{\frac{2}{3}+\frac{m-1}{l}} P_{y}^{\frac{2}{3}+\frac{2+\delta}{3}\cdot\frac{n-1}{r}} +gP_{x}^{\frac{3}{l}(1-m)}P_{y}^{\frac{\varepsilon}{r}(1-n)} \bigg], \label{hamk}
\end{align}
where
\begin{align}
f&=\frac{3}{2}\bigg[\frac{(\rho lr)^2}{6V_{j0}}\bigg]^{1/3}\bigg(\frac{-x}{\rho m} \bigg)^{\frac{2}{3}-\frac{1}{l}}\bigg(\frac{-y}{ n}\bigg)^{\frac{2}{3}-\frac{2+\delta}{3r}}, \label{fxy} \\
g&=V_0\bigg(\frac{-x}{\rho m} \bigg)^{\frac{3}{l}}\bigg(\frac{-y}{ n}\bigg)^{\frac{\varepsilon}{r}}. \label{gxy}
\end{align}
From \ref{hamk}, we can see $\tilde{H}$ is a constrained Hamiltonian system, since $P_z=p_N=0$ implies that $0=\dot{P}_z=-\partial\tilde{H}/\partial z$. Thus,
\begin{equation}\label{constrk}
P_{x}^{\frac{2}{3}+4\frac{m-1}{l}}P_{y}^{\frac{2}{3}+ \frac{2+\delta+3\varepsilon}{3}\cdot\frac{n-1}{r}}=\frac{g}{f}\equiv\lambda.
\end{equation}
For simplicity, we can choose
\begin{equation}
l=6\qquad\mbox{and}\qquad r=\textstyle\frac{1}{2}(2+\delta+3\varepsilon),
\end{equation}
so that $\lambda$ is a positive constant:
\begin{equation}\label{eqwdwprovtc}
\lambda=\frac{2V_0}{3}\bigg[\frac{V_{j0}}{6(r\rho)^2} \bigg]^{1/3}.
\end{equation}
Now, if we require that quantization gives a second order partial differential equation and that \ref{funcger} is not degenerate, we must choose $m=n=3/2$. Then constraint \ref{constrk} becomes
\begin{equation}
P_xP_y=\lambda.
\end{equation}
Now canonical quantization $\hat{P}_{j}=-i\hbar\partial_j$ can directly be applied, leading to the Wheeler-DeWitt equation
\begin{equation}\label{wdweq}
\frac{\partial^2\psi}{\partial x\partial y}=-\frac{\lambda}{\hbar^2}\psi,
\end{equation}
where $\psi(x,y)$ is the stationary wave function of the Universe. The basic solution is the plane wave
\begin{equation}\label{wdweqpsik}
\psi_k(x,y)=e^{i(kx+\omega y)/\hbar},
\end{equation}
where $k\neq0$ is a real constant and $\omega\equiv\lambda/k$. Let us now briefly review the core ideas of Bohmian interpretation of quantum mechanics to apply them to \ref{wdweq}.

\subsection{Bohmian Interpretation}\label{intdbb}
As an answer to the incompleteness  of quantum mechanics claimed by A. Einstein, B. Podolsky, and N. Rosen in \cite{epr}, some authors argued in favour of standard interpretation, like N. Bohr \cite{PhysRev.48.696} and L. E. Ballentine \cite{RevModPhys.42.358} later. However, this criticism inspired also an alternative interpretation of quantum mechanics, suggested by D. Bohm in \cite{PhysRev.85.166,PhysRev.85.180}. Bohmian mechanics provides a method to associate a deterministic dynamics for an individual quantum system, thus avoiding the incompleteness pointed out in \cite{epr}. To illustrate those ideas, consider Schr\"odinger equation for a single particle
\begin{equation}\label{schr}
-\frac{\hbar^2}{2m}\nabla^2\psi+V(\mathbf{x})\psi=i\hbar\frac{\partial\psi}{\partial t},
\end{equation}
where $\psi(\mathbf{x},t)$ is the wave function and $V(\mathbf{x})$ is a potential. Since $\psi$ is complex, it can be written as $\psi=Re^{iS/\hbar}$, where $R$ and $S$ are real functions. Thus, the imaginary and the real parts of \ref{schr} become, respectively
\begin{subequations}
	\label{eqscr}
\begin{align}
\frac{\partial R^2}{\partial t}+\nabla\cdot\bigg(\frac{R^2\nabla S}{m}\bigg) &= 0, \label{schrim}\\
\frac{\partial S}{\partial t} +\frac{|\nabla S|^2}{2m}+V(\mathbf{x})+Q(\mathbf{x})&=0, \label{schrre}
\end{align}
\end{subequations}
where
\begin{equation}\label{potqschr}
Q(\mathbf{x})=-\frac{\hbar^2}{2m}\frac{\nabla^2R}{R}.
\end{equation}
Equation \ref{schrim} is a continuity equation. As for \ref{schrre}, except for the term $Q$, it is a Hamilton-Jacobi equation with $S$ playing the role of the Hamilton principal function. D. Bohm suggested in \cite{PhysRev.85.166,PhysRev.85.180} to interpret that as follows: the quantum $\nabla S$ can be associated with the classical momentum of the particle by
\begin{equation}\label{eqguiaschr1p}
\mathbf{p}=\nabla S=\hbar\img\frac{\nabla\psi}{\psi},
\end{equation}
in analogy with Hamilton-Jacobi formalism, and the additional term $Q$ is understood as being a quantum contribution (of order $\hbar^2$) to the total amount of energy. Because of that, $Q$ is called the {\it quantum potential}. Now, since $\mathbf{p}=m\mathbf{\dot{x}}$, it follows that \ref{eqguiaschr1p} gives a method to obtain deterministic trajectories for the particle. Thus, for each solution $\psi$, there is a whole family of possible trajectories. That is why $\psi$ is sometimes called the {\it pilot wave} that guides the solution through the trajectories and \ref{eqguiaschr1p} is called the {\it guidance equation}. In standard quantum mechanics, the recovery of classical dynamics follows from the correspondence principle \cite{PhysRev.48.696}. In Bohmian quantum mechanics, it follows from the quantum Hamilton-Jacobi equation \ref{schrre} that the classical mechanics is recovered when $Q=0$.

It can be shown that Bohmian interpretation can describe all basic numerical features of standard quantum mechanics \cite{cushing2013bohmian,durr2009bohmian,holland_1993}. Further discussions and applications can be found in \cite{freire2014quantum,pladevall2019applied,HOLLAND199395}. As mentioned in the introduction, there are some conceptual arguments in favor of Bohmian mechanics in quantum cosmology. In the references \cite{PintoNeto:2004uf,PINTO-NETO2000,PINTONETO2000194,PhysRevD.95.123522,PhysRevD.97.083517}, it is shown how to generalize the above Bohmian formalism to models of quantum gravity and quantum cosmology. Among other things, they show how that formalism exhibits quantum effects, but also describes scalar and tensor perturbations in spacetime. We will now apply that formalism to \ref{wdweq}.

In what follows, the comma denotes partial derivative. Let us write $\psi(x,y)=R(x,y)e^{iS(x,y)/\hbar}$, where $R$ and $S$ are real functions. Thus, the imaginary and real parts of \ref{wdweq} are, respectively,
\begin{subequations}
	\label{reimwdw}
\begin{align}
RS_{,xy}+ R_{,x}S_{,y}+R_{,y}S_{,x}&=0, \label{imgwdw}\\
-S_{,x}S_{,y}+\lambda+\frac{\hbar^2}{R} R_{,xy}&=0, \label{rewdw}
\end{align}
\end{subequations}
where we have set $N=z=1$ (cosmic time). Equation \ref{imgwdw} is the analogous of the continuity equation \ref{schrim}. Now, rearranging \ref{rewdw} to compare it with classical stationary Hamilton-Jacobi equation for $\tilde{H}$, the quantum potential is given by
\begin{equation}\label{pqqtc}
	Q=\hbar^2f S_{,x}^{-\frac{1}{4}} S_{,y}^{-\frac{\varepsilon}{2r}}\frac{R_{,xy}}{R},
\end{equation}
and the guidance equations are
\begin{equation}\label{eqguiatc}
	P_x=S_{,x},\qquad\mbox{and}\qquad P_y=S_{,y}.
\end{equation}

\subsection{Recovering Classical Solutions}
For the plane wave \ref{wdweqpsik}, $R=1$ and $S=kx+\omega y$, so the quantum potential \ref{pqqtc} vanishes. Therefore, in analogy with the Bohmian interpretation for Schr\"odinger equation, we expect that for a null quantum potential the classical solutions are recovered. If that is the case, we can say the quantum formalism here developed is consistent. In fact, from \ref{canctransf} and \ref{eqguiatc}, we can recover the quantum values of the momenta, given by guidance equations
\begin{align}
	p_a &=-6\rho k^{3/2}a^5, \label{paqtc} \\
	p_{\phi} &=-r(\lambda/k)^{3/2}\phi^{r-1}.
\end{align}
Then, from those quantum relations and from \ref{afip}, we obtain the following system:
\begin{subequations}
	\label{sistq}
\begin{align}
\dot{a}&=\frac{V_0}{9\rho k^{3/2}}\frac{\phi^{\varepsilon}}{a^2}, \label{sistdinqtca}\\
\dot{\phi}&=\frac{2V_0}{3r}\bigg(\frac{k}{\lambda}\bigg)^{3/2}a^3 \phi^{-\frac{1}{2}(\delta+\varepsilon)}. \label{sistdinqtcf}
\end{align}
\end{subequations}
Hence, setting
\begin{align}
\rho=1/9k^{3/2},
\end{align}
the quantum system \ref{sistq} becomes entirely equivalent to classical system \ref{eqelc3}, for power law potentials \ref{potlptc}, for any powers $\delta,\varepsilon$. In other words, for the solution \ref{wdweqpsik}, that gives a null quantum potential, the classical equations are recovered, as was expected. This result can be extended to the case where both $V$ and $V_j$ are exponentials, by defining $\varphi\equiv e^{\phi}$ in \ref{ljmse} and adapting all calculations, as we mentioned above. Thus, we can say, in particular, that classical solutions are recovered in the classical limit of the Bohmian formalism for power law \ref{afifp} and de Sitter \ref{afids} solutions. Therefore, Bohmian interpretation can be successfully applied for those cases of \ref{lagtot}. In physical terms, we proved that the quantum model is consistent for de Sitter, matter-dominated, stiff matter-dominated, and radiation-dominated solutions for the scale factor.

\section{Conclusions}\label{conc}
In this letter, we have explored some aspects of the background cosmology of Fab Four John theory \ref{lagtot}. Due to its structure, the dynamics is governed by the first-order system \ref{eqelc3}, from which we found a big variety of cosmological solutions, including basic phases of the evolution of the universe, such as accelerated expansion, radiation-dominated, and matter-dominated eras. We also have shown that bouncing and cyclic universes are possible in this theory. All those solutions follow from the structure of the potential $V(\phi)$ and from the scalar field interpretation as a time scale. This last result is a direct consequence of \ref{eqelc3}.

For that derivation, we have set $\phi=t$ for simplicity, but the scalar field can be any differentiable increasing function defined on real line. Thus, $\phi$ may, in principle, represent any strictly increasing physical quantity. In that case, different choices must be made for $V$ and $V_j$, in order to keep all solutions above. Thanks to the simple structure of \ref{eqelc3}, this is always possible, if the diffeomorphism condition is still satisfied by $\phi$. We have to stress that this letter is intended to be a background analysis, so further questions concerning perturbations are still a matter of investigation for future works.

We have also presented a preliminary quantum approach to Fab Four John. Because of the odd structure of the Hamiltonian, the quantization is not straightforward. The John kinetic term is proportional to $(p_ap_{\phi})^{2/3}$, thus a canonical transformation must be performed. But we have shown as a first result that, at least for power law and exponential functions $V,V_j$, the quantization is well behaved, in the sense that the classical solutions are recovered when the quantum potential vanishes.

In conclusion, we can say that the big variety of solutions for the nonminimal derivative coupling $L_{john}$ here studied raises some questions. What should be the cosmological solutions if the coupling constant $\kappa$ in  \ref{rpg5mv} is replaced by the general function $V_j(\phi)$? Is it possible to obtain such results when that coupling is only a contribution? Since  \ref{rpg5mv} do not contradicts the gravitational waves constraint, those are important questions to investigate in future works.

\section*{Acknowledgments}
We thank to Ingrid Ferreira da Costa, Felipe de Melo Santos, Jorge Zanelli, Joseph Buchbinder, Nelson Pinto Neto, and Sergey Sushkov for very important discussions about this paper. This study was financed in part by the \emph{Coordena\c{c}\~ao de Aperfei\c{c}oamento de Pessoal de N\'ivel Superior} - Brazil (CAPES) - Finance Code 001 and also by FAPES and CNPq from Brazil. OFP thanks the Alexander von Humboldt foundation for funding and the Institute for Theoretical Physics of the Heidelberg University for kind hospitality.

\bibliography{mybibfile}
\end{document}